\documentstyle[12pt]{article}

\catcode`\@=11
\long\def\@makefntext#1{
\protect\noindent \hbox to 3.2pt {\hskip-.9pt
$^{{\ninerm\@thefnmark}}$\hfil}#1\hfill}		%CAN BE USED

\def\@makefnmark{\hbox to 0pt{$^{\@thefnmark}$\hss}}  %ORIGINAL

\def\ps@myheadings{\let\@mkboth\@gobbletwo
\def\@oddhead{\hbox{}
\rightmark\hfil\ninerm\thepage}
\def\@oddfoot{}\def\@evenhead{\ninerm\thepage\hfil
\leftmark\hbox{}}\def\@evenfoot{}
\def\sectionmark##1{}\def\subsectionmark##1{}}

\renewcommand{\thefootnote}{\fnsymbol{footnote}}

\newcounter{sectionc}\newcounter{subsectionc}\newcounter{subsubsectionc}
\renewcommand{\section}[1] {\vspace*{0.5cm}\addtocounter{sectionc}{1}
\setcounter{subsectionc}{0}\setcounter{subsubsectionc}{0}\noindent
	{\normalsize\bf\thesectionc. #1}\par\vspace*{0.4cm}}
\renewcommand{\subsection}[1] {\vspace*{0.6cm}\addtocounter{subsectionc}{1}
	\setcounter{subsubsectionc}{0}\noindent
	{\normalsize\it\thesectionc.\thesubsectionc. #1}\par\vspace*{0.4cm}}
\renewcommand{\subsubsection}[1]
{\vspace*{0.6cm}\addtocounter{subsubsectionc}{1}
	\noindent {\normalsize\rm\thesectionc.\thesubsectionc.\thesubsubsectionc.
	#1}\par\vspace*{0.4cm}}

\newcounter{appendixc}
\newcounter{subappendixc}[appendixc]
\newcounter{subsubappendixc}[subappendixc]

\renewcommand{\appendix}[1] {\vspace*{0.6cm}
        \refstepcounter{appendixc}
        \setcounter{figure}{0}
        \setcounter{table}{0}
        \setcounter{equation}{0}
        \renewcommand{\thefigure}{\Alph{appendixc}.\arabic{figure}}
        \renewcommand{\thetable}{\Alph{appendixc}.\arabic{table}}
        \renewcommand{\theappendixc}{\Alph{appendixc}}
        \renewcommand{\theequation}{\Alph{appendixc}.\arabic{equation}}
        \noindent{\bf Appendix \theappendixc #1}\par\vspace*{0.4cm}}

\def\abstracts#1{{

\centering{\begin{minipage}{12.2truecm}\footnotesize\baselineskip=12pt\noindent
	\centerline{\footnotesize ABSTRACT}\vspace*{0.3cm}
	\parindent=0pt #1
	\end{minipage}}\par}}

\renewenvironment{thebibliography}[1]
	{\begin{list}{\arabic{enumi}.}
	{\usecounter{enumi}\setlength{\parsep}{0pt}
\setlength{\leftmargin 1.25cm}{\rightmargin 0pt}
	 \setlength{\itemsep}{0pt} \settowidth
	{\labelwidth}{#1.}\sloppy}}{\end{list}}

\newcounter{itemlistc}
\newcounter{romanlistc}
\newcounter{alphlistc}
\newcounter{arabiclistc}

\newcommand{\fcaption}[1]{
        \refstepcounter{figure}
        \setbox\@tempboxa = \hbox{\footnotesize Fig.~\thefigure. #1}
        \ifdim \wd\@tempboxa > 6in
           {\begin{center}
        \parbox{6in}{\footnotesize\baselineskip=12pt Fig.~\thefigure. #1}
            \end{center}}
        \else
             {\begin{center}
             {\footnotesize Fig.~\thefigure. #1}
              \end{center}}
        \fi}

\newcommand{\tcaption}[1]{
        \refstepcounter{table}
        \setbox\@tempboxa = \hbox{\footnotesize Table~\thetable. #1}
        \ifdim \wd\@tempboxa > 6in
           {\begin{center}
        \parbox{6in}{\footnotesize\baselineskip=12pt Table~\thetable. #1}
            \end{center}}
        \else
             {\begin{center}
             {\footnotesize Table~\thetable. #1}
              \end{center}}
        \fi}

\def\@citex[#1]#2{\if@filesw\immediate\write\@auxout
	{\string\citation{#2}}\fi
\def\@citea{}\@cite{\@for\@citeb:=#2\do
	{\@citea\def\@citea{,}\@ifundefined
	{b@\@citeb}{{\bf ?}\@warning
	{Citation `\@citeb' on page \thepage \space undefined}}
	{\csname b@\@citeb\endcsname}}}{#1}}

\newif\if@cghi
\def\cite{\@cghitrue\@ifnextchar [{\@tempswatrue
	\@citex}{\@tempswafalse\@citex[]}}
\def\citelow{\@cghifalse\@ifnextchar [{\@tempswatrue
	\@citex}{\@tempswafalse\@citex[]}}
\def\@cite#1#2{{$\null^{#1}$\if@tempswa\typeout
	{IJCGA warning: optional citation argument
	ignored: `#2'} \fi}}

 1
 1
 1

\font\ninerm=cmr9

\begin{document}

\newcommand{\st}{\scriptstyle}
\newcommand{\sst}{\scriptscriptstyle}
\newcommand{\mco}{\multicolumn}
\newcommand{\epp}{\epsilon^{\prime}}
\newcommand{\vep}{\varepsilon}
\newcommand{\ra}{\rightarrow}
\newcommand{\ppg}{\pi^+\pi^-\gamma}
\newcommand{\vp}{{\bf p}}
\newcommand{\ko}{K^0}
\newcommand{\kb}{\bar{K^0}}
\newcommand{\al}{\alpha}
\newcommand{\ab}{\bar{\alpha}}
\def\be{\begin{equation}}
\def\ee{\end{equation}}
\def\bea{\begin{eqnarray}}
\def\eea{\end{eqnarray}}
\def\CPbar{\hbox{{\rm CP}\hskip-1.80em{/}}}
%temp replacement due to no font
%%%%%%%%%%%%%%%
% My Macros   %
%%%%%%%%%%%%%%%
%   \newcommand{\cZ}{\cal{Z}}
%   \newtheorem{def}{Definition}[section]
%   ...
\newcommand{\spa}[3]{\left\langle#1\,#3\right\rangle}
\newcommand{\spb}[3]{\left[#1\,#3\right]}
\newcommand{\LP}{\left(}
\newcommand{\RP}{\right)}
\newcommand{\LB}{\left[}
\newcommand{\RB}{\right]}
\newcommand{\Tr}{\mathop{\rm Tr}\nolimits}
\newcommand{\e}{\epsilon}
\newcommand{\cg}{c_\Gamma}
\newcommand{\hf}{\textstyle{1\over2}}
\newcommand{\Li}{\mathop{\rm Li}\nolimits}
\newcommand{\Ls}{\mathop{\rm Ls}\nolimits}
\newcommand{\Ll}{\mathop{\rm L}\nolimits}
\newcommand{\gluino}{{\tilde g}}
\newcommand{\qb}{{\bar q}}
\newcommand{\susy}{{\rm SUSY}}
\newcommand{\tree}{{\rm tree}}
\newcommand{\oneloop}{{\rm 1-loop}}
\newcommand{\Atree}{A^{\rm tree}}
\newcommand{\Atreestar}{A^{\rm tree\,*}}
\newcommand{\Aloop}{A^{\rm 1-loop}}
\newcommand{\Aloopstar}{A^{\rm 1-loop\,*}}
\newcommand{\pol}{\varepsilon}
\newcommand{\si}{\sigma}
\newcommand{\ns}{n_{\mskip-2mu s}}
\newcommand{\nf}{n_{\mskip-2mu f}}
\newcommand{\ib}{{\bar\imath}}
%

%%%%%%%%%%%%%%%%%%%%%%%
% Beginning of text   %
%%%%%%%%%%%%%%%%%%%%%%%

\vspace{-1.1in}
\noindent
\hfill SLAC--PUB--95--6771

\noindent
\hfill hep-ph/9503261

\vspace{0.10in}

\centerline{\normalsize\bf EFFICIENT ANALYTIC COMPUTATION OF}
\baselineskip=15pt
\centerline{\normalsize\bf HIGHER-ORDER QCD AMPLITUDES\footnote{%
Research supported by the Department of Energy under grants
DE-FG03-91ER40662 and DE-AC03-76SF00515, by the Alfred P. Sloan
Foundation under grant BR-3222, by the National Science Foundation
under grant PHY-9218990, by the {\it Commissariat \`a l'Energie
Atomique\/} of France, and by NATO Collaborative Research Grants
CRG--921322 and CRG--910285.  Talk presented by L.D.}}

%\vfill
%\vspace*{0.6cm}
\centerline{\footnotesize ZVI BERN, GORDON CHALMERS}
\baselineskip=13pt
\centerline{\footnotesize\it Department of Physics, UCLA,
 Los Angeles, CA 90024, USA}
\centerline{\footnotesize E-mail: bern@physics.ucla.edu}
\vspace*{0.3cm}
\centerline{\footnotesize LANCE DIXON}
\baselineskip=13pt
\centerline{\footnotesize\it Stanford Linear Accelerator Center,
  Stanford University, Stanford, CA 94309, USA}
\centerline{\footnotesize E-mail: lance@slac.stanford.edu}
\vspace*{0.3cm}
\centerline{\footnotesize DAVID C. DUNBAR}
\baselineskip=13pt
\centerline{\footnotesize\it University College of Swansea, UK}
\centerline{\footnotesize E-mail: D.C.Dunbar@swansea.ac.uk}
\vspace*{0.3cm}
\centerline{\footnotesize and}
\vspace*{0.3cm}
\centerline{\footnotesize DAVID A. KOSOWER}
\baselineskip=13pt
\centerline{\footnotesize\it Service de Physique Th\'eorique,
  Centre d'Etudes de Saclay,
  F-91191 Gif-sur-Yvette cedex, France}
\centerline{\footnotesize E-mail: kosower@amoco.saclay.cea.fr}

%\vfill
\vspace*{0.7cm}
\abstracts{
We review techniques simplifying the analytic calculation
of one-loop QCD amplitudes with many external legs, for use in
next-to-leading-order corrections to multi-jet processes.
Particularly useful are the constraints imposed by perturbative
unitarity, collinear singularities and a supersymmetry-inspired
organization of helicity amplitudes.
Certain sequences of one-loop helicity amplitudes with an
arbitrary number of external gluons have been obtained using
these constraints.
}

%\vspace*{0.5cm}
\normalsize\baselineskip=14pt
\setcounter{footnote}{0}
\renewcommand{\thefootnote}{\alph{footnote}}

\section{Total Quantum-number Management}
The calculation of one-loop QCD amplitudes with many external
quarks and gluons is a bottleneck that must be navigated in order
to obtain next-to-leading-order (NLO) corrections to multijet
processes, for precision comparison with collider experiments.
The full correction has a real (bremsstrahlung) part as well as
a virtual part.
Efficient techniques for computing the tree amplitudes entering the
real part have been available for several
years\cite{ManganoParke};
however, significant numerical work is required to combine these
parts into a finite answer.
In this talk we ignore the numerical subtleties\footnote{%
Such subtleties have recently been discussed for the
energy-energy correlation in $e^+e^-$ annihilation\cite{GSCE}.},
and focus on techniques for computing analytically the one-loop
amplitudes entering the virtual part.

In principle it is straightforward to compute one-loop amplitudes
by drawing all Feynman diagrams and evaluating them using standard
reduction techniques for the loop integrals.  In practice this method
becomes extremely inefficient and cumbersome as the number of
external legs grows, because there are:
\par\noindent
1. {\bf too many diagrams} --- many diagrams are related by gauge
invariance, and
\par\noindent
2. {\bf too many terms in each diagram} --- nonabelian gauge boson
self-interactions are complicated.
\par\noindent
Consequently, intermediate expressions tend to be vastly more
complicated than the final results, when the latter are
represented in an appropriate way.

A useful organizational framework, that helps tame the size of
intermediate expressions, is Total Quantum-number Management
(TQM), which suggests to:
\par\noindent
$\bullet$ Keep track of all possible information about external
particles --- namely, {\it helicity} and {\it color} information.
\par\noindent
$\bullet$ Keep track of quantum {\it phases} by computing the
transition amplitude rather than the cross-section.
\par\noindent
$\bullet$ Use the helicity/color information to decompose the amplitude
into simpler, gauge-invariant pieces, called {\it sub-amplitudes}
or {\it partial amplitudes}.
\par\noindent
$\bullet$ Square amplitudes to get probabilities, and sum over
helicities and colors to obtain unpolarized cross-sections,
only at the very {\it end} of the calculation.
\par\noindent
Carrying out the last step explicitly would generate a large
analytic expression; however, at this stage one would
typically make the transition to numerical evaluation,
in order to combine the virtual and real corrections.
The use of TQM is hardly new, particularly in tree-level
applications\cite{ManganoParke} ---
but it is especially useful at loop level.

As an example, consider the one-loop amplitude for $n$ external
gluons, all taken to be outgoing.
We generalize the $SU(3)$ color group to
$SU(N_c)$, and label the gluons $i=1,2,\ldots,n$ by their adjoint
color indices $a_i=1,2,\ldots,N_c^2-1$, and helicities
$\lambda_i = \pm$.  The helicity decomposition uses gluon circular
polarization vectors expressed in terms of
massless Weyl spinors\cite{SpinorHelicity}.  The color
decomposition\cite{LoopColor} is performed in
terms of traces of $SU(N_c)$ generators $T^a$ in the
fundamental representation, with $\Tr(T^aT^b)=\delta^{ab}$,
\begin{eqnarray*}
&&\hskip -8mm
 {\cal A}^\oneloop_n \LP \{k_i,\lambda_i,a_i\}\RP =
  g^n\Biggl[
    \sum_{\sigma \in S_n/Z_n}
    N_c\,\Tr\LP T^{a_{\sigma(1)}}\cdots T^{a_{\sigma(n)}}\RP\
     A_{n;1}(\sigma(1^{\lambda_1}),\ldots,\sigma(n^{\lambda_n}))
     \nonumber \\
&& \hskip -8mm
 +\ \sum_{c=2}^{\lfloor{n/2}\rfloor+1}
      \sum_{\sigma \in S_n/S_{n;c}}
    \Tr\LP T^{a_{\sigma(1)}}\cdots T^{a_{\sigma(c-1)}}\RP\
    \Tr\LP T^{a_{\sigma(c)}}\cdots T^{a_{\sigma(n)}}\RP\
     \ A_{n;c}(\sigma(1^{\lambda_1}),\ldots,\sigma(n^{\lambda_n}))
     \Biggr]\, ,
\end{eqnarray*}
where $A_{n;c}$ are the partial amplitudes,
$g$ is the gauge coupling, $S_n$ is the set of all permutations of
$n$ objects, while $Z_n$ and $S_{n;c}$ are the subsets of $S_n$
that leave the corresponding single and double trace structures
invariant.

The $A_{n;1}$ are more basic, and are called {\it primitive
amplitudes}, because:
\par\noindent
{\it a}. They only receive contributions from diagrams with a
particular cyclic ordering of the gluons around the loop,
which greatly simplifies their analytic structure.
\par\noindent
{\it b}. The remaining $A_{n;c>1}$ can be
generated\cite{LoopColor,SusyFour} as sums of permutations
of the $A_{n;1}$.\footnote{%
For amplitudes with external quarks as well as gluons, the
primitive amplitudes are not a subset of the partial amplitudes;
new color-ordered objects have to be defined\cite{qqggg}.}

Even the $A_{n;1}$ are not all
independent, due to parity and cyclic invariance.
For example, for $n=5$ only four are independent,
$A_{5;1}(1^+,2^+,3^+,4^+,5^+)$,
$A_{5;1}(1^-,2^+,3^+,4^+,5^+)$,
$A_{5;1}(1^-,2^-,3^+,4^+,5^+)$, and
$A_{5;1}(1^-,2^+,3^-,4^+,5^+)$.
The first two of these are not required at NLO
because the corresponding tree helicity amplitudes vanish, and are
very simple for the same reason.  For $n_f$ quark flavors, they
are given by\cite{FiveGluon}
\begin{eqnarray*}
&& \hskip -8mm
  A_{5;1}(1^+,2^+,3^+,4^+,5^+) = {iC\over 48\pi^2}
  {  \spa1.2\spb1.2\spa2.3\spb2.3 + \spa4.5\spb4.5\spa5.1\spb5.1
   + \spa2.3\spa4.5\spb2.5\spb3.4
 \over \spa1.2 \spa2.3 \spa3.4 \spa4.5 \spa5.1 }\ , \nonumber \\
&& \hskip -8mm
  A_{5;1}(1^-,2^+,3^+,4^+,5^+) = {iC\over 48\pi^2}
  {1\over\spa3.4^2 }
 \Biggl[-{ \spb2.5^3 \over \spb1.2\spb5.1 }
 + { \spa1.4^3\spb4.5\spa3.5 \over \spa1.2\spa2.3\spa4.5^2 }
 - { \spa1.3^3\spb3.2\spa4.2 \over \spa1.5\spa5.4\spa3.2^2 } \Biggr]
 \ ,
\end{eqnarray*}
where $C=1-{n_f\over N_c}$ and $\spa{j}.l$, $\spb{j}.l$ are spinor
inner products\cite{SpinorHelicity,ManganoParke}.
Analytic expressions for the other two primitive amplitudes
are more complex but still ``fit on a page''\cite{FiveGluon}.
In contrast, the color- and helicity-summed virtual correction to the
cross-section, built from permutation sums of the two primitive
amplitudes, would fill hundreds of pages.

\section{Analytic Properties (and Supersymmetry)}
There are at least five different ways to compute the
partial/primitive amplitudes:
\par\noindent
1. Traditional Feynman diagrams (in the helicity, color-ordered basis).
\par\noindent
2. Rules derived from superstring theory\cite{StringBased}.
\par\noindent
3. Rules inspired by superstring theory\cite{StringInspired}.
\par\noindent
4. Recursive construction\cite{Mahlon} (see also the talk by
Mahlon\cite{MahlonBSMIV} in these proceedings).
\par\noindent
5. Exploitation of their analytic properties (and supersymmetry).
\par\noindent
Here we just discuss route 5, which can be the most efficient route
to the answer.

The analytic behavior of loop amplitudes includes both cuts and poles.
Since primitive amplitudes are ``color-ordered'' (property {\it a}),
they have cuts and poles only in channels formed by the sum of
{\it cyclically adjacent} momenta, $(k_i+\cdots+k_{i+r-1})^2$.
In a (massless) supersymmetric theory, because of the improved
ultraviolet behavior, the cuts alone are enough to reconstruct the
full one-loop amplitude\cite{SusyOne}.
The cuts are computable in closed form in many cases.
For example, the infinite sequence of maximal helicity-violating
amplitudes in $N=4$ super-Yang-Mills theory are
given\cite{SusyFour} (in $4-2\e$ dimensions, through ${\cal O}(\e^0)$)
by a sum of known scalar box integral functions $F_{n:r;i}$
($j$ and $k$ are the only gluons with negative helicity):
\begin{displaymath}
  A_{n;1}^{N=4}(1^+,\ldots,j^-,\ldots,k^-,\ldots,n^+) =
 i\, {(4\pi\mu^2)^\e\over 16\pi^2}
  {\Gamma(1+\e) \Gamma^2(1-\e) \over \Gamma(1-2\e)}
   { \spa{j}.k^4 \over \spa1.2\spa2.3\cdots\spa{n}.1 }\, V_n\,,
\end{displaymath}
where
\begin{eqnarray*}
  V_{2m+1} &=& \sum_{r=2}^{m-1} \sum_{i=1}^n F_{n:r;i}^{2{\rm m}\,e}
                              + \sum_{i=1}^n F_{n:i}^{1{\rm m}}\ ,
  \nonumber \\
    V_{2m} &=& \sum_{r=2}^{m-2} \sum_{i=1}^n F_{n:r;i}^{2{\rm m}\,e}
                              + \sum_{i=1}^n F_{n:i}^{1{\rm m}}
                        + \sum_{i=1}^{n/2} F_{n:m-1;i}^{2{\rm m}\,e}\ .
\end{eqnarray*}

Supersymmetric results can be used to trade QCD calculations
with internal gluons for somewhat easier calculations where scalars
replace the gluons.   For an amplitude with all external gluons,
we rewrite the internal gluon loop $g$ (and fermion loop $f$) as a
supersymmetric contribution plus a complex scalar loop $s$,
\begin{eqnarray*}
  g &=& (g+4f+3s)\ -\ 4(f+s)\ +\ s\ =\ [N=4]\ -\ 4\,[N=1]\ +\ s, \\
  f &=& (f+s)\ -\ s\ =\ [N=1]\ -\ s,
\end{eqnarray*}
where $[N=1]$ represents the contribution of an $N=1$ chiral
supermultiplet.  The scalar contribution cannot be reconstructed
directly from its cuts because of an additive ``polynomial''
ambiguity.
It seems possible to fix this ambiguity by inspecting the
factorization (pole) behavior of the amplitude, namely the
limits where two (or more) adjacent momenta become collinear.
The general form of these limits for one-loop amplitudes has recently
been proven\cite{BernChalmers}.
For the special case of identical gluon helicities, $(1^+,\ldots,n^+)$,
the limits were successfully used to construct an
ansatz\cite{AllPlus} which was subsequently confirmed by recursive
techniques\cite{Mahlon}.
If one can show that polynomial ambiguities for arbitrary
helicity configurations can be determined uniquely and efficiently
from factorization limits, then one would have a
general technique for constructing one-loop QCD amplitudes
without ever evaluating genuine loop diagrams.


\section{References}
\begin{thebibliography}{9}
\bibitem{GSCE} E.W.N. Glover and M.R. Sutton, Phys. Lett. B342:375
         (1995);
         K. Clay and S. Ellis, hep-ph/9502223.
\bibitem{ManganoParke} M. Mangano and S. Parke, Phys. Rep. 6:301
         (1991).
\bibitem{SpinorHelicity} F.A. Berends, R. Kleiss, P. De Causmaecker,
         R. Gastmans and T.T. Wu, Phys. Lett. B103:124 (1981);
         P. De Causmaeker, R. Gastmans,  W. Troost and T.T. Wu,
         Nucl. Phys. B206:53 (1982);
         R. Kleiss and W.J. Stirling, Nucl. Phys. B262:235 (1985);
         J.F. Gunion and Z. Kunszt, Phys. Lett. B161:333 (1985);
         Z. Xu, D.-H.\ Zhang and L. Chang, Nucl. Phys. B291:392 (1987).
\bibitem{LoopColor}  Z. Bern and D. Kosower, Nucl. Phys. B362:389
         (1991).
\bibitem{SusyFour} Z. Bern, D. Dunbar, L. Dixon and D. Kosower,
         Nucl. Phys. B425:217 (1994).
\bibitem{qqggg} Z. Bern, L. Dixon and D. Kosower,
         Nucl. Phys. B437:259 (1995).
\bibitem{FiveGluon}  Z. Bern, L. Dixon and D. Kosower, Phys. Rev. Lett.
         70:2677 (1993).
\bibitem{StringBased} Z. Bern and D. Kosower, Phys. Rev. Lett. 66:1669
         (1991); Nucl. Phys. B379:451 (1992).
\bibitem{StringInspired} Z. Bern and D. Dunbar, Nucl. Phys. B379:562
         (1992).
\bibitem{Mahlon} G. Mahlon, Phys. Rev. D49:2197 (1994);
         Phys. Rev. D49:4438 (1994).
\bibitem{MahlonBSMIV} G. Mahlon, hep-ph/9412350, in these proceedings.
\bibitem{BerendsGiele} F.A. Berends and W.T. Giele, Nucl. Phys.
         B306:759 (1988).
\bibitem{SusyOne} Z. Bern, D. Dunbar, L. Dixon and D. Kosower,
         Nucl. Phys. B435:59 (1995).
\bibitem{AllPlus} Z. Bern, G. Chalmers, L. Dixon and D. Kosower,
         Phys. Rev. Lett. 72:2134 (1994).
\bibitem{BernChalmers} Z. Bern and G. Chalmers, hep-ph/9503236.
\end{thebibliography}
\end{document}